\documentclass[%
 reprint,
 amsmath,amssymb,
 aps,
]{revtex4-1}

\usepackage{subfig}
\usepackage{graphicx}% Include figure files
\usepackage{dcolumn}% Align table columns on decimal point
\usepackage{bm}% bold math
\usepackage{color}

\begin{document}

\preprint{APS/123-QED}

\title{Tunable free energy and heat flux between two-dimensional materials}

\author{Hamidreza Simchi}
\email{simchi@alumni.iust.ac.ir}
\affiliation {Department of Physics, Iran University of Science and Technology, Narmak, Tehran 16844, Iran} \affiliation{Semiconductor Technoloy Center, P.O.Box 19575-199, Tehran, Iran}%Lines break automatically or can be forced with \\

\date{\today}% It is always \today, today,

\begin{abstract}
We study the free energy across a stratified media made by graphene $(G)$ and/or molybdenum disulfide $(MoS_{2})$. The flux depends not only on the number of $G/MoS_{2}$ layers but also on the priority of graphene layer respect to $MoS_{2}$ layer in the media. The rule is; the thinner layer should be the nearest neighborhood of leftmost and rightmost sides of the media for getting the highest energy flux. The free energy can be tuned by applying an external gate voltage to $MoS_{2}$ layers due to the tunable property of the dielectric constant of $MoS_{2}$ by the gate voltage. Also we show in the $silicon/MoS_{2}/silicon$ three-body configuration, the photon heat tunneling is amplified significantly due to the increasing the number of the coupled modes. Due to the amplifying effect, this mechanism could be exploited to improving detection ability in the near infrared detection systems. 
\end{abstract}

\maketitle

%\tableofcontents

\section{Introduction}
A semiconductor photovoltaic (PV) cell absorbs photons and generates hole-electron pairs. All photons, with energy less than the energy gap $(E_{g})$ of semiconductor, are only dissipated into heat within the atomic lattice and do not contribute to the pair generation. Some parts of photons with energy larger than $E_{g}$ are converted to electron-hole pairs but some parts of their energy are  dissipated via phonon excitation. A body at temperature $T$ radiates electromagnetic waves due to the thermal and quantum fluctuations\cite{R1,R2,R3,R4}. By defining the thermal wavelength as $\lambda_{th}=\hbar c/(k_{B}T)$, we can consider two different regimes for heat flow. At distances smaller than $\lambda_{th}$ the evanescent term of heat flow dominates (near-field) and for distances larger than $\lambda_{th}$, the propagating radiation term dominates (far-field) \cite{R1,R2,R3,R4,R5,R6}. Hence, the near-field thermal radiation confined on the surface is potentially important source of energy. Now if we consider a PV cell in the proximity of a thermal emitter, the energy can be extracted when the evanescent photons tunnel toward the cell. Such device has been proposed less than twenty years ago and is called near-field thermophotovoltaic (NTPV) system \cite{R7}. Therefore, If we can amplify the photon heat tunneling, more energy will be extracetd by NTPV. 
\\
It has been shown that, the near-field heat transfer between two plates varies as $d^{-2}$ where $d$ is vacuum gap between the plates \cite{R2,R8,R9,Rten}. It should be noted that the $d^{-2}$ dependence is for contribution from p-polarized electromagnetic waves only since the contribution from s-polarized waves will asymptotically reach a constant as $d\rightarrow 0$\cite{R11,R12}. For heavily doped semiconductors, the contribution of p-polarized waves dominates over that of s-polarized waves when $d<10$ {\color {red} $(nm)$}\cite{R13} but for metals, the crossover can take place at much shorter distances \cite{R14,R15}. It has been shown that there are optimal values of real and imaginary parts of the dielectric constants of plates, for which, the near-field heat transfer is maximized\cite{R16}. Messina et al., have introduced the concept of three-body amplification of heat exchanged at nanoscale between two media\cite{R17}. We have shown  how one can overcome the limitations of Messina's model\cite{R17} by using graphene as intermediate layer\cite{R18}.Of course, the theory of van der Waals forces can be generalized to different multilayer geometries and a closed form solution that explicitly depend on the number of layers in the multilayer system can be derived\cite{R19,R20,R35}.
\\
In this paper, we consider a multilayer structure of graphene and/or $MoS_{2}$ and study the free energy across the multilayer. We show how the energy depends not only on the number of layers but also on the priority of graphene layer respect to $MoS_{2}$ layer in the stratified media. There is a rule. The thinner layer should be the nearest neighborhood of leftmost and rightmost sides of stratified media for getting the highest interaction energy.  Since the dielectric constant of $MoS_{2}$ can be tuned by applying an external gate voltage $(V_{g})$, we show how one can tune the free energy across the multilayer structure by applying $(V_{g})$. In addition, for showing the probable application of heat flow, we consider a three-body configuration as $silicon/MoS_{2}/silicon$, and show how amplifying the heat flow directly implies an amplification of the sensitivity of this configuration 
when we add the intermediate $MoS_{2}$ layer. The structure of article is as follows: In section II, the calculation model is explained. In section III, the numerical calculation, results, and discussion are provided. The summary is presented in section IV.
\section{Calculation method}
In this section, we discuss about the dielectric constant of $MoS_{2}$, and graphene, firstly. Then, we study the free energy in multilayer structures and derive the closed form of free energy density in terms of the number of layers. Finally, we review the heat flow amplification in three-body configuration shortly, based on our previous work \cite{R18}. 
\subsection{Dielectric constant of $MoS_{2}$}
A large range of values for dielectric constant $(\varepsilon)$ of $MoS_{2}$ has been found in experiments \cite{R21,R22,R23,R24,R25}. It has been shown that $\varepsilon$ can be manipulated by an external electric field\cite{R26}. Mukherjee et al., have studied the complex electrical permittivity of monolayer $MoS_{2}$ in the near and visible range\cite{R27}. They have shown that the real and imaginary part of $\varepsilon$ can be written as\cite{R27}:
\begin{equation}
\varepsilon_{R}=\varepsilon_{\infty}+{\frac{137}{\pi}}\sum_{j=0}^{5}{\frac{a_{j}\omega_{P}^{2}}{(\omega_{j}^{2}-\omega^{2})^2+\omega^2b_{j}^{2}}}(\omega_{j}^{2}-\omega^2)
\end{equation}
\begin{equation}
\varepsilon_{I}={\frac{137}{\pi}}\sum_{j=0}^{5}{\frac{a_{j}\omega_{P}^{2}}{(\omega_{j}^{2}-\omega^{2})^2+\omega^2b_{j}^{2}}}(\omega^2b_{j})+\alpha exp(-{\frac{\omega-\mu}{2\sigma^{2}}})
\end{equation}
where, $\omega_{P}=28.3\times 10^{-3}$ electron volt $(eV)$, $\varepsilon_{\infty}=4.44$, $\alpha=23.224$,$\mu=2.7723$$(eV)$,$\sigma=0.3089$$(eV)$\cite{R27}. Other constants of above equations are given in table I by considering $a=a_{j}/\hbar (eV)$, $b=b_{j}/\hbar (eV)$ and $c=\omega_{j}/\hbar (eV)$\cite{R27}.
\begin{table}[b]%The best place to locate the table environment is directly after its first reference in text
\caption{\label{tab:table1}%
The values of constants of Eqs. 1 and 2.}
\begin{ruledtabular}
\begin{tabular}{lcdr}
\textrm{j}&
\textrm{a}&
\multicolumn{1}{c}{\textrm{b}}&
\textrm{c}\\
\colrule
0 & 2.0089E5 & 1.0853E-2 & 0\\
1 & 5.7534E4 & 5.9099E-2 & 1.88\\
2 & 8.1496E4 & 1.1302E-1 & 2.03\\
3 & 8.2293E4 & 1.1957E-1 & 2.78\\
4 & 3.3130E5 & 2.8322E-1 & 2.91\\
5 & 4.3906E6 & 7.8515E-1 & 4.31\\
\end{tabular}
\end{ruledtabular}
\end{table}
However, by applying a gate voltage $V_{g}$, the Fermi energy $E_{F}$ changes as\cite{R27}
\begin{equation}
E_{F}=\hbar^{2}\pi C V_{g}/0.7m_{e}e^{3}      
\end{equation}
where, $C=1.2\times 10^{-8}$ $(F cm^{-2})$, $m_{e}$ and $e$ are mass and charge of electron, respectively. Since, the dependency of constants $a_{j}$ and $b_{j}$ to $E_{F}$ is as follows\cite{R27}:

\begin{equation}
a_{j}^{\prime}=a_{j}exp(-12\pi(E_{F}-k_{B}T)^{2})          
\end{equation}
\begin{equation}
b_{j}^{\prime}=b_{j}/exp(-12\pi(E_{F}-k_{B}T)^{2})           
\end{equation}
One can tune the dielectric constant of $MoS_{2}$ by applying a gate voltage.
\subsection{Dielectric constant of graphene}
It has been shown that for $q/k_{F}>\omega/E_{F}$, the dielectric constant of graphene can be written as\cite{R18,R28}:
\begin{equation}
\varepsilon(q,i\omega)=1+{\frac{\pi g c q}{1096\varepsilon_{m}v\sqrt{q^2-(i\omega/v)^2}}}
\end{equation}
where, $\varepsilon_{m}$ is the average dielectric constant of the surrounding media, $g=4$,$c=3\times 10^8$$m/s$, $v\approx 10^6 m/s$,$k_{F}=\sqrt{4pi\rho/g}$ and $\rho$ is the average electron density \cite{R18,R28}.

\subsection{ Free energy}

Fig.1 shows a multilayer slab composed of $N+1$ layers $A $ and $N$ layers of $B$ between the semi-infinite region $(L)$ on the left-hand side and the semi-infinite region $(R)$ on the right-hand side.

\begin{figure}
\includegraphics{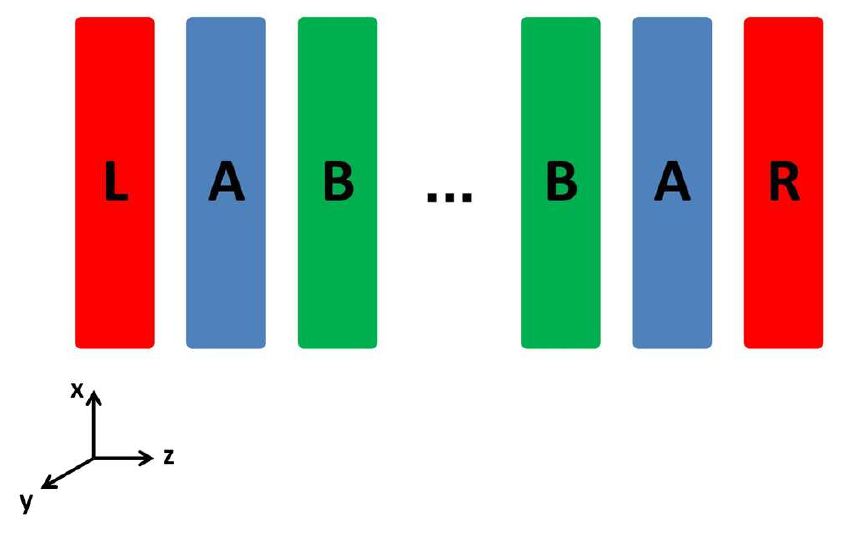}% Here is how to import EPS art
\caption{\label{fig:epsart} (Color online) A multilayer slab composed of $N+1$ layers $A $ and $N$ layers of $B$.}
\end{figure}
We assume that the system is homogeneous in $(x,y)$ plane and the electric and magnetic fields have the below form:
\begin{equation}
\vec{E}=\vec{e}(z)e^{i\vec{q}.\vec{\rho}}
\end{equation}
\begin{equation}
\vec{H}=\vec{h}(z)e^{i\vec{q}.\vec{\rho}}
\end{equation}
where, $\vec{\rho}=(x,y)$ and $\vec{q}=(q_{x},q_{y})$. In each dielectric medium $i$, the function $\vec{e}_{i}(z)$ must satisfy the Helmholtz equation
\begin{equation}
{\frac{d^2\vec{e}_{i}}{dz^2}}+(\omega^2\varepsilon_{i}\mu_{i}/c^2-q^2)\vec{e}_{i}=0
\end{equation}
whose solution has the form
\begin{equation}
\vec{e}_{i}=\vec{A}_{i}e^{\rho_{i}z}+\vec{B}_{i}e^{-\rho_{i}z}        
\end{equation}
with
\begin{equation}
\rho_{i}^2=q^2-{\frac{\varepsilon_{i}\mu_{i}\omega^2}{c^2}}
\end{equation}
At interface between two neighborhood layers $(i-1)$ and $(i)$, the transverse components of  $\vec{E}$ i.e., $E_{x}$ and $E_{y}$, are continuous while in the longitudinal direction it is the displacement $\vec{D}$ that is continuous i.e.,
 \begin{equation}
E_{i-1,x}=E_{i,x}, E_{i-1,y}=E_{i,y}      
\end{equation} 
\begin{equation}
 D_{i-1,z}=D_{i,z}     
\end{equation}
The same holds also for $\vec{B}$ and $\vec{H}$.
It should be noted that since both fields are divergence free, the spatial components of $\vec{A_{i}}$ and $\vec{B_{i}}$ satisfy the below equations:
\begin{equation}
A_{i,z}=-{\frac{i}{\rho_{i}}}(q_{x}A_{i,x}+ q_{y}A_{i,y})    
\end{equation}
\begin{equation}
B_{i,z}={\frac{i}{\rho_{i}}}(q_{x}B_{i,x}+ q_{y}B_{i,y})     
\end{equation}
which, we use the below identity
\begin{equation}
\nabla . (\psi \vec{a})=\vec{a} . \nabla \psi+\psi \nabla . \vec{a}
\end{equation}
By using Eqs. 12, and 13 and also Eqs. 14, and 15, it can be shown that:
\begin{widetext}
\begin{equation}
A_{i,z} e^{\rho_{i}z}+ B_{i,z} e^{-\rho_{i}z}=\frac{\varepsilon_{i-1}}{\varepsilon_{i}}(\frac{-i}{\rho_{i-1}})(q_{x}A_{i-1,x}+ q_{y}A_{i-1,y}) e^{\rho_{i-1}z}+
\frac{\varepsilon_{i-1}}{\varepsilon_{i}}(\frac{i}{\rho_{i-1}})(q_{x}B_{i-1,x}+ q_{y}B_{i-1,y}) e^{-\rho_{i-1}z}
\label{eq:wideeq}
\end{equation}
\end{widetext}
Also, by using Eq. 12, it can be shown that:
\begin{widetext}
\begin {equation}
{-\frac{\rho_{i}}{i}}A_{i,z}e^{\rho_{i}z}+{\frac{\rho_{i}}{i}}B_{i,z}e^{\rho_{i}z}=(q_{x}A_{i-1,x}+q_{y}A_{i-1,y})e^{\rho_{i-1}z}+(q_{x}B_{i-1,x}+q_{y}B_{i-1,y})e^{-\rho_{i-1}z}
\label{eq:wideeq}
\end{equation}
\end{widetext}
Therefore,
\begin{widetext}
\begin{equation}
B_{i,z}={\frac{\varepsilon_{i}\rho_{i-1}+\varepsilon_{i-1}\rho_{i}}{2\rho_{i}\varepsilon_{i}}}[({\frac{\rho_{i}\varepsilon_{i-1}-\varepsilon_{i}\rho_{i-1}}{\varepsilon_{i}\rho_{i-1}+\rho_{i}\varepsilon_{i-1}}})A_{i-1,z} e^{\rho_{i-1}z_{i-1}+\rho_{i}z_{i}}+B_{i-1,z}e^{\rho_{i}z{i}-\rho_{i-1}z_{i-1}}]
\label{eq:wideeq}
\end{equation}
\end{widetext}
and
\begin{widetext}
\begin{equation}
A_{i,z}={\frac{\varepsilon_{i}\rho_{i-1}+\varepsilon_{i-1}\rho_{i}}{2\rho_{i}\varepsilon_{i}}}[({\frac{\rho_{i}\varepsilon_{i-1}-\varepsilon_{i}\rho_{i-1}}{\varepsilon_{i}\rho_{i-1}-\rho_{i}\varepsilon_{i-1}}})B_{i-1,z}e^{\rho_{i-1}z_{i-1}-\rho_{i}z_{i}}+A_{i-1,z} e^{-\rho_{i}z_{i}-\rho_{i-1}z_{i}}]
\label{eq:wideeq}
\end{equation}
\end{widetext}
By using Eqs. 19 and 20 it can be shown that\cite{R19}
\begin{widetext}
\begin{equation}
\begin{pmatrix}
A_{i,z}\\B_{i,z}
\end{pmatrix}
=
\begin{pmatrix}
{1}&{-\Delta_{i-1,i} e^{-2\rho_{i-1}d_{i-1}}}\\
-\Delta_{i-1,i}&{e^{-2\rho_{i-1}d_{i-1}}}
\end{pmatrix}
\begin{pmatrix}
A_{i-1,z}\\B_{i-1,z}
\end{pmatrix}
\end{equation}
\end{widetext}
where, $d_{i-1}=l_{i-1,i}-l_{i-2,i-1}$ and $l_{i}$ is the position of the discontinuity the ith and the $i-1 th$ layers. Also,
\begin{equation}
\Delta_{i-1,i}={\frac{\varepsilon_{i-1}\rho_{i}-\varepsilon_{i}\rho_{i-1}}{\varepsilon_{i-1}\rho_{i}+\varepsilon_{i}\rho_{i-1}}}
\end{equation}
The first matrix in the right-hand side of Eq. 21, called M, can be decomposed to two separate matrices called discontinuity matrix, $D_{i,i-1}$ and propagator matrix, $T_{i-1}$ such that $M=D\times T$ where: 
\begin{equation}
D_{i,i-1}=
\begin{pmatrix}
{1}&{-\Delta_{i,i-1} }\\
-\Delta_{i,i-1}&{1}
\end{pmatrix}
\end{equation}
and
\begin{equation}
T_{i-1}=
\begin{pmatrix}
{1}&{0}\\
{0}&{e^{-2\rho_{i}d_{i-1}}}
\end{pmatrix}
\end{equation}
Now, we can generalize this relation to any number of layers. We start at the leftmost layer $L$ and ending at the rightmost layer $R$ with $N$ layers in between and write\cite{R19}:
\begin{widetext}
\begin{equation}
M=D_{R,N-1}\times T_{N-1}\times D_{N-1,N-2}\times T_{N-2}\times D_{N-2,N-3}\times T_{N-3}\times \cdot\cdot\cdot\times T_{1}\times D_{1,L}
\label{eq:wideeq}
\end{equation}
\end{widetext}
If in a periodic structure $a_{R}^{T}=(A_{R},B_{R})$ , $a_{L}^{T}=(A_{L},B_{L})$, and $a_{R}^{T}=M a_{L}^{T}$, then
\begin {equation}
A_{R}=m_{11}A_{L}+m_{12}B_{L}
\end{equation}
\begin {equation}
B_{R}=m_{21}A_{L}+m_{22}B_{L}
\end{equation}
But for incoming wave $B_{L}=0$ and for outgoing wave $A_{R}=0$ (see Eq.10). Therefore, $m_{11}A_{L}=0$ and $B_{R}=m_{21}A_{L}$. Since, $A_{L}\neq 0$ then $m_{11}=0$. The only excitations of the fields that satisfy the boundary conditions are those obtained from solving the secular equation $m_{11}=0$\cite{R19}. Moreover, in Lifshitz theory the fluctuation free energy is directly related to the $m_{11}$ and is written as\cite{R19}
\begin{widetext}
\begin{equation}
F(N,a,b)=k_{B}T\sum_{q}\sum_{n=0}^{\infty,\prime} ln(m_{11}^{TE})+ k_{B}T\sum_{q}\sum_{n=0}^{\infty,\prime} ln(m_{11}^{TM})
\label{eq:wideeq}
\end{equation}
\end{widetext}
where, $k_{B}$ is Boltzmann constant and the primed sum signifies that the $n=0$ term is taken with the weight $1/2$\cite{R19}. Sarlah et al., have shown that in a nonisotropic homogeneous uniaxial slab composed of periodic layers, $F(N,a,b)$ can be written approximately as\cite{R29}:
\begin{widetext}
\begin{equation}
F(N,a,b)={\frac{k_{B}T}{2\pi}}\sum_{n=0}^{\infty,\prime}\int_{0}^{\infty}  q\ dq\ log (1+\Delta_{RA}\Delta_{AL} e^{-2\rho_{A}(a+N(a+b)})   
\label{eq:wideeq}
\end{equation}
\end{widetext}
where,
\begin{equation}
\Delta_{RA}=\frac{(\rho_{A}\varepsilon_{A}-\rho_{A}\varepsilon_{\perp})}
{(\rho_{A}\varepsilon_{A}+\rho_{A}\varepsilon_{\perp})}
=-\Delta_{AL}
\label{eq:wideeq}
\end{equation}
and
\begin{equation}
\rho_{A}^{2}={\frac{\varepsilon_{\perp}}{\varepsilon_{\parallel}}}(q^2-{\frac{\varepsilon_{\parallel}\omega^2}{c^2}})
\end{equation}

\begin{equation}
\varepsilon_{\perp}={\frac{1}{a+b}}(a\varepsilon_{A}+b\varepsilon_{B}) 
\end{equation}

\begin{equation}
\varepsilon_{\parallel}={\frac{1}{a+b}}(a/\varepsilon_{A}+b/\varepsilon_{B})
\end{equation}

We will use the Eq.29 for doing the numerical calculations. Finally, the dependency of interaction energy to the number of $(AB)$ layers, $(N)$,is given by Hamaker coefficient $H(N,a,b)$ as\cite{R19}

\begin {equation}
H(N,a,b)=12\pi (a+b)^{2} F(N,a,b)/N^{2}
\end{equation} 
Electromagnetic fluctuations are related to one of the most fundamental phenomena in nature, namely Brownian motion. The fluctuations of the electromagnetic field determine a large class of important physical phenomena, such as the Casimir force and radiative heat transfer. In the other words, The electromagnetic fluctuations are the cornerstone of the Casimir physics. These fluctuations are described in the fluctuational electrodynamics by introducing a random field into the Maxwell equation. Although the Casimir force arises from electromagnetic fluctuations, real photons are not involved. Quantum mechanically, these fluctuations can be described in terms of virtual photons of energy that are equal to the zero-point energies of the electromagnetic modes of the system. 
In addition, the propagating electromagnetic waves always exist outside any body due to thermal and quantum fluctuations of the current density inside the body. Quantum fluctuations are related to the uncertainty principle, and exist also at zero temperature. Thermal fluctuations are due to the irregular thermal motion of the particles in the medium, and vanish at zero temperature. The electromagnetic field created by this fluctuating current density exists not only in the form of propagating
waves but also in the form of evanescent waves, which are damped exponentially with the distance away from the surface of the body. This fluctuating electromagnetic field exists even at zero temperature, generated by quantum fluctuations\cite{R36}. Therefore, we encounter with two phenomena i.e., there are two evanescent waves inside the gap ($d$) between two parallel plates when $d<10$  $(nm)$.One component comes from free energy and other component comes from heat transfer. 

It has been shown that the evanescent photons can tunnel between the nearest neighborhood layers\cite{R18}. In above, we described the  evanescent waves related to the free energy and in next section we will review  the  evanescent waves related to the heat transfer. Anyhow, by decreasing (increasing) the evanescent photons the less (more) photons exist between the nearest neighborhod layers. In the other words, if we can adjust the evanescent photons we will able to adjust the photon density between the nearest neighborhood layers. 

\subsection{Three-body photon heat amplification}
Fig.2 shows a three-body configuration including silicon as leftmost and rightmost regions and a monolayer of $MoS_{2}$ as intermediate layer. It has been shown that the TM evanescent part $P_{ev}$ of the heat flow between two bodies is equal to\cite{R17,R18}:
\begin{widetext}
\begin{equation}
P_{ev}={\frac{\hbar}{\pi^2 d^2}}\int_{0}^{\infty}\omega d\omega(n_{B}(E,T_{1})-n_{B}(E,T_{2}))\times\int_{0}^{\infty}\gamma d\gamma {\frac{Im(\varepsilon_{1}) Im(\varepsilon_{2})}{(\mid(\varepsilon_{1}+1)(\varepsilon_{2}+1)-(\varepsilon_{1}-1)(\varepsilon_{2}-1) e^{-\gamma} \mid)^2}}e^{-\gamma}
\label{eq:wideeq}
\end{equation}
\end{widetext}
 where, $d$ is vacuum distance between two bodies, $\gamma=-2p\omega d/c$, $\varepsilon_{1}$ and $\varepsilon_{2}$ are dielectric constant, $c$ is  velocity of light, and $n_B(E,T_{i})=[e^{\hbar\omega/{K_{B}T}}-1]^{-1}$. The relation of $p=\mid \hat p\mid$ is given in our previous work\cite{R18}.  If the $MoS_{2}$ layer is absent, then\cite{R18}:
\begin{widetext}
\begin{equation}
X_{1}\equiv\int_{0}^{\infty}\gamma d\gamma {\frac{Im(\varepsilon_{1}) Im(\varepsilon_{2})}{(\mid(\varepsilon_{1}+1)(\varepsilon_{2}+1)-(\varepsilon_{1}-1)(\varepsilon_{2}-1) e^{-\gamma} \mid)^2}}e^{-\gamma}=\int_{0}^{\infty}\gamma d\gamma {\frac{Im(\varepsilon_{1}) Im(\varepsilon_{2})}{(\mid(\varepsilon_{Si}+1)^2-(\varepsilon_{Si}-1)^2 e^{-\gamma} \mid)^2}}e^{-\gamma}
\label{eq:wideeq}
\end{equation}
\end{widetext}

where, $X=(n_{B}(E,T_{1})-n_{B}(E,T_{2}))\times X_{1}$ is the monochromatic radiative heat flux in the near-field. If in cylindrical coordinates, the wave vector is written as $k_{j}=\beta\vec{r}+\alpha _{j}\vec{z}$ with $k_{j}^2=\beta^2+\alpha^2$, then the cut off wave vector parallel to the surface is $\beta_{c}=\pi/d_{c}$ where $d_{c}$ is the lattice constant. The value of $X$ is normalized by $\beta_{c}^2/8$ i.e., $X^*=8X/\beta_{c}^2$\cite{R17,R18}. After adding the intermediate $MoS_{2}$ layer, its temperature is chosen such that the total heat flux on this layer to be zero\cite{R18}. We will use the Eqs.35 and 36 for calculating the heat transfer in three-body configuration, shown in Fig,2.  Of course, it is assumed that a vaccum gap exist between $Si$ and $MoS_{2}$ and in consequence two dielectric constants only appear in Eq.35\cite{R18}. 

\begin{figure}
\includegraphics{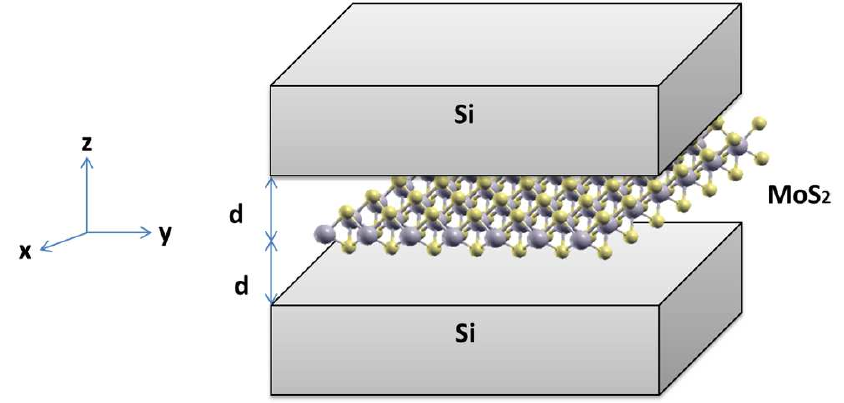}% Here is how to import EPS art
\caption{\label{fig:epsart}  (Color online) Three-body configuration including silicon as leftmost and rightmost regions and a monolayer of $MoS_{2}$ as intermediate layer.}
\end{figure}
\section{Result and discussion}
\subsection{Without gate voltage}
Let us start with the dielectric constant of $MoS_{2}$. Fig.3(a) shows the dielectric constant, $\varepsilon$, of monolayer $MoS_{2}$ via wavelength. Four peaks at $\lambda=293, 451, 619$, and $666$$nm$ are seen which is in good agreement with published data\cite{R27}. Now, we consider a multilayer structure of graphene and $MoS_{2}$ and study the photon tunneling across the structure in absence of gate voltage. First, we assume that, the layer $A$ and layer $B$ of Fig.1  are made by graphene ($MoS_{2}$,)and $MoS_{2}$(graphene), respectively. Fig.4(a)  shows the interaction free energy across the structure via the number of layers, $N$.

\begin{figure*}
\includegraphics[width=0.4\linewidth]{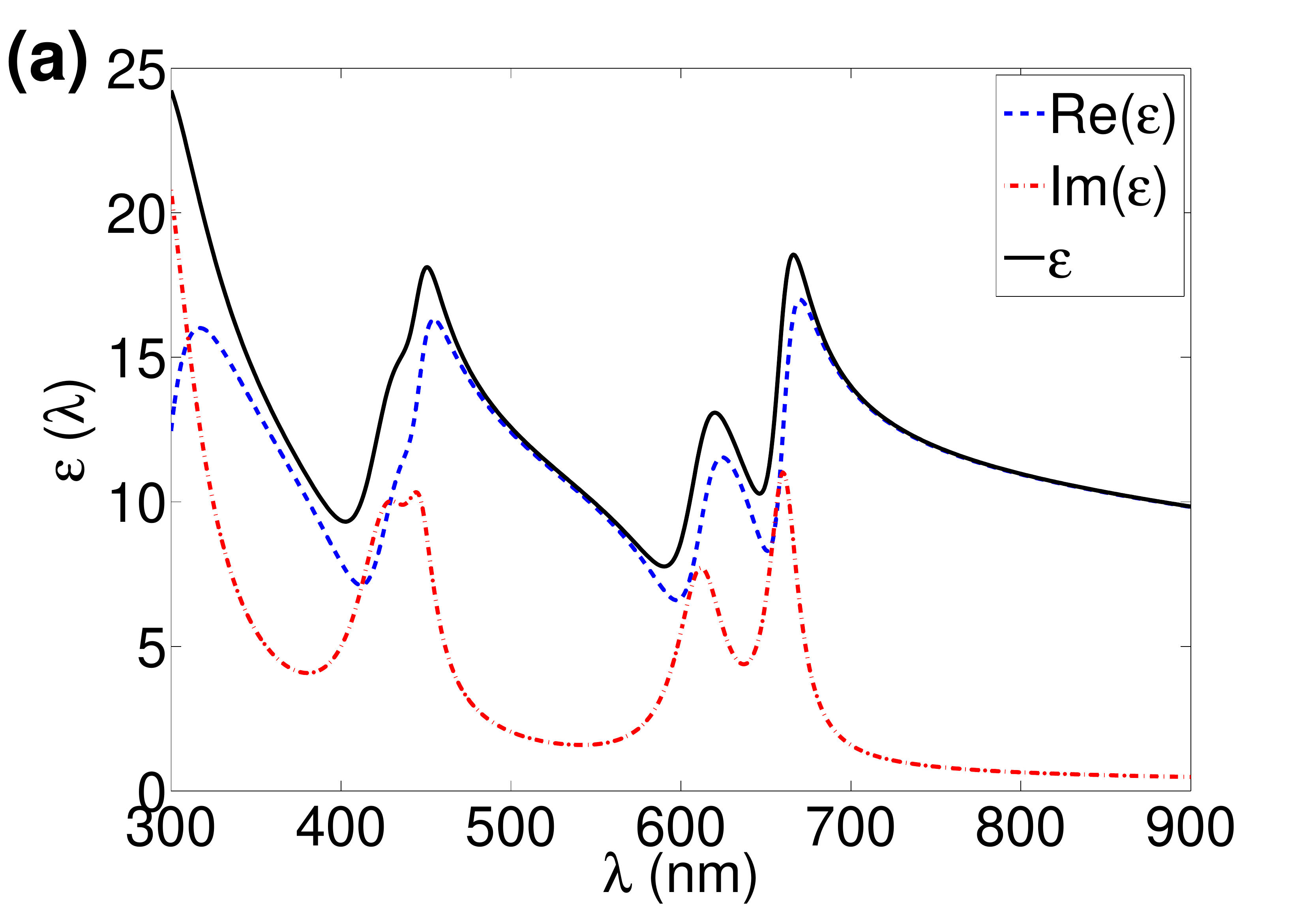}
\hspace{20pt}
\includegraphics[width=0.44\linewidth]{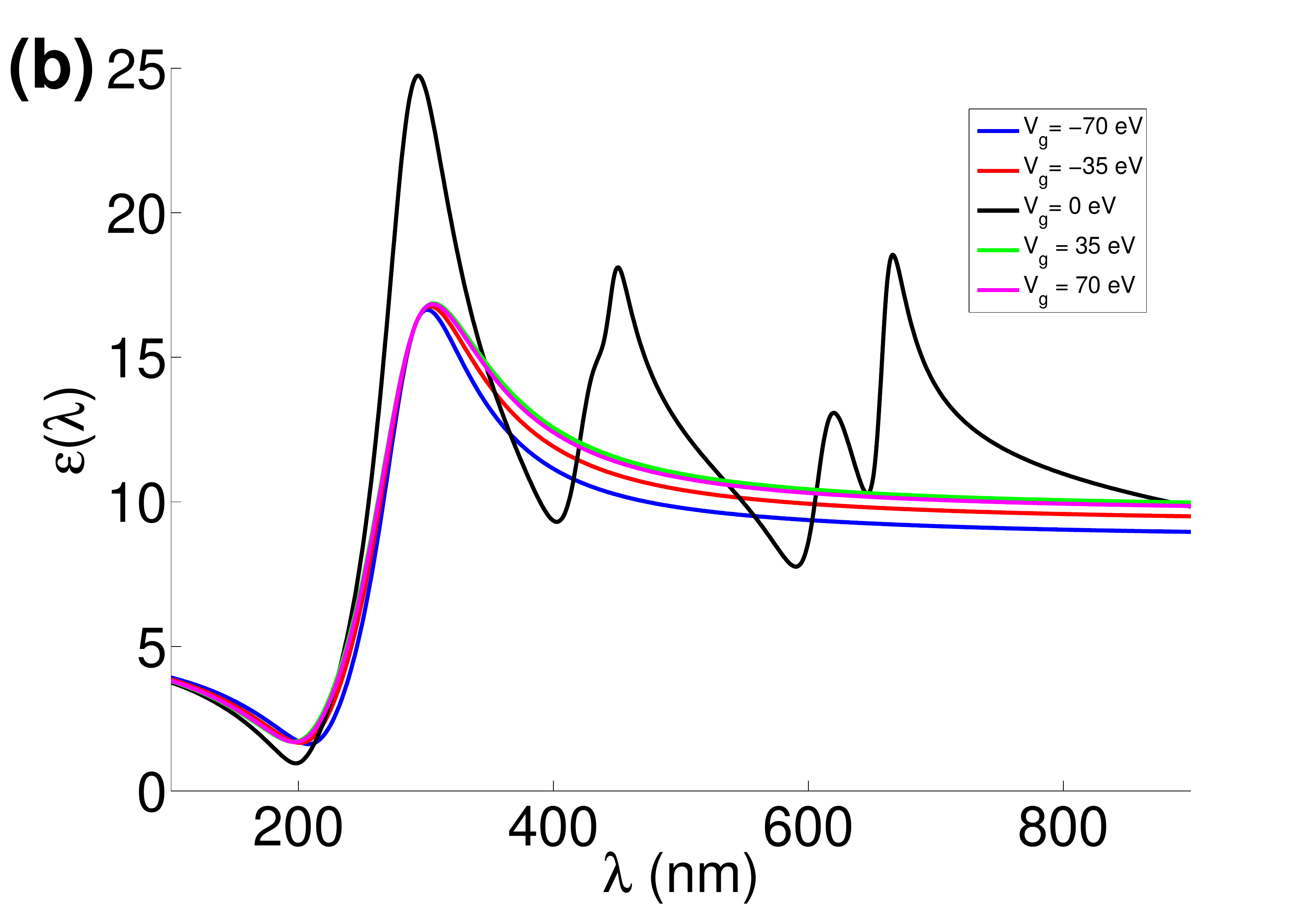}
\caption{(Color online) (a) The dielectric constant of $ (MoS_{2})$ when $V_{g}=0 (eV)$ and (b) $V_{g}\neq 0$. }
\label{fig:Fig3}
\end{figure*} 
 As it is expected, the free energy decreases by increasing the number of layers. It is well known that the $TE (TM)$ mode of oscillations decays with factor $e^{-\gamma}$ from surface of the body\cite{R17,R18}; and therefore, by increasing the distance between leftmost side  and rightmost side, the free energy decreases. Also, the exponential term of Eq.29 depends on thickness of A-layer as $(N+1)a$ and thickness of B-layer as $Nb$. The thickness of  graphene and $MoS_{2}$ is 1 Angstrom $(A^0)$ and 3.1 $A^0$, respectively\cite{R28,R32}. Therefore, when the first layer of stratified structure is $MoS_{2}$ the power of exponential term of Eq.29 is bigger than that of graphene. Then, as Fig.4(a) shows, the free energy of $G/MoS_{2}/G$ is higher than $MoS_{2}/G/MoS_{2}$. Therefore, there is a general rule. The thinner layer should be the nearest neighborhood of leftmost and rightmost sides of stratified media for getting the highest interaction energy. Fig.4(b) shows the Hamaker coefficient via $N$. The Hamaker coefficient $H(N,a,b)$ is defined by Eq.34.
As it shows, the transition to the continuum approximation begins from $N=10$ for $G/MoS_{2}/G$ structure while it begins from $N=5$ for $MoS_{2}/G/MoS_{2}$ structure. Thus, from this point of view, the $MoS_{2}/G/MoS_{2}$ structure is better than $G/MoS_{2}/G$ structure.
\begin{figure}
\includegraphics[width=0.8\linewidth]{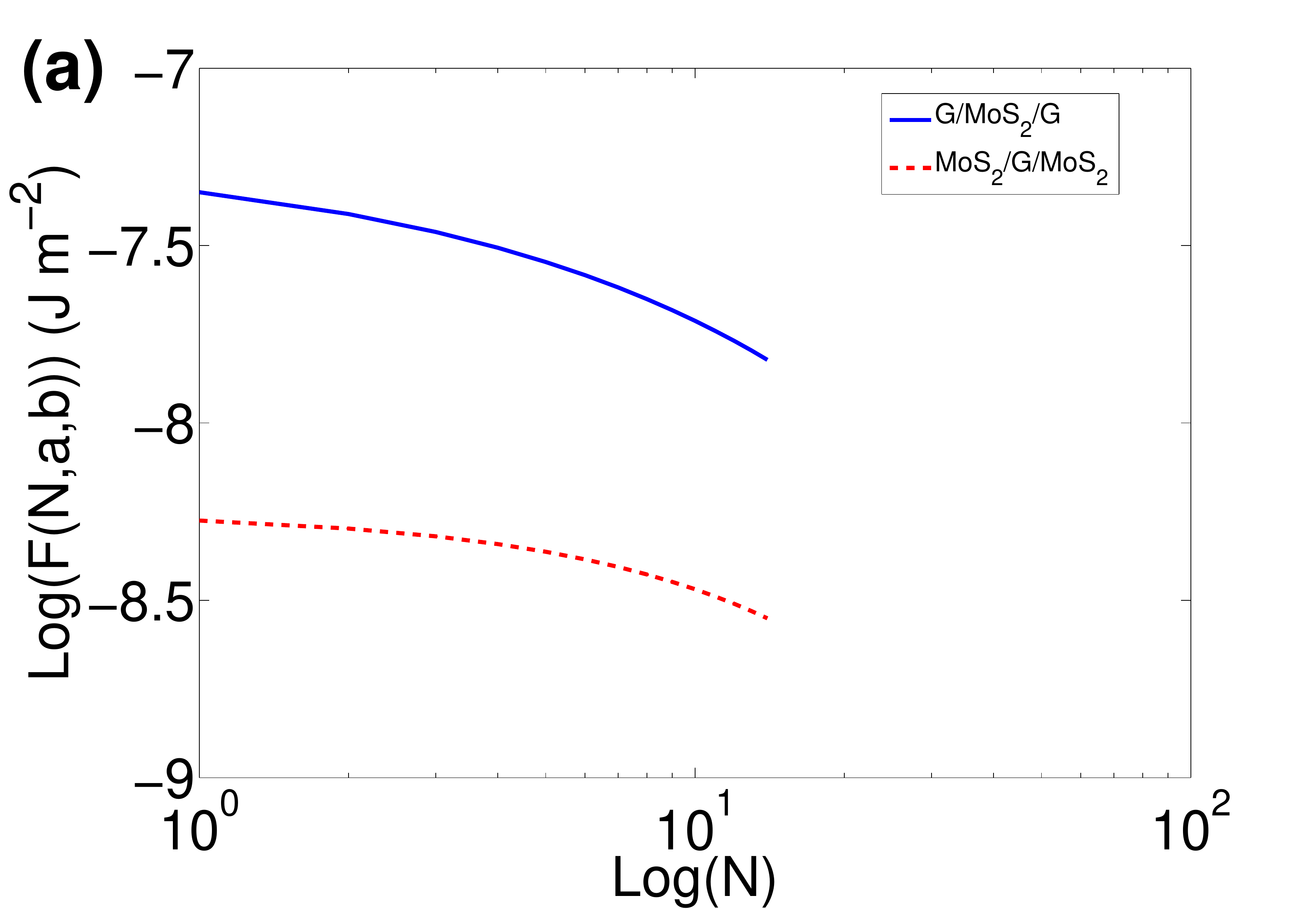}
\hspace{20pt}
\includegraphics[width=0.8\linewidth]{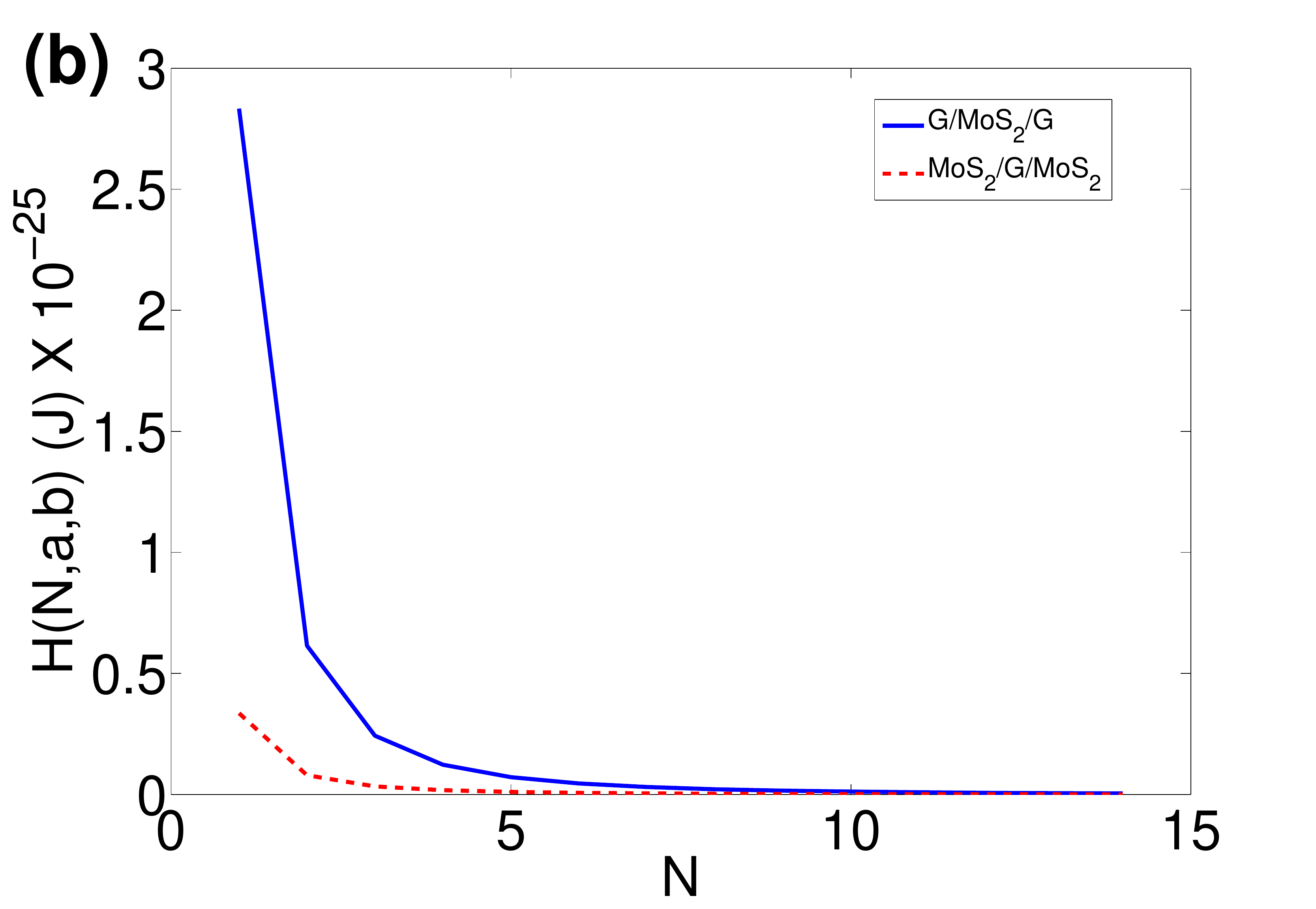}
\caption{(Color online) (a) The interaction free energy across $N$ layers of $MoS_{2}$ (graphene) and $N+1$ layers of graphene$ (MoS_{2})$. The thickness of graphene and $MoS_{2}$ is equal to 1 and 3.1 Angstrom $(A^0)$. (b) Hamaker coefficient via number of layers. }
\label{fig:Fig4}
\end{figure}

\subsection{With gate voltage}
It has been shown that, the Fermi energy changes under applying the gate voltage $(V_{g})$, and in consequence, the absorption spectrum changes too\cite{R30,R31}. It means that, $\varepsilon(\lambda)$, changes if an external gate voltage is applied. Fig. 3(b) shows, how $\varepsilon(\lambda)$ changes via $V_{g}$. As it shows, the height of first peak decreases but the other peaks disappear when $V_{g}$ is applied. Since, the heat photon tunneling between two bodies depends on the dielectric constant of bodies\cite{R17,R18} this property of monolayer $MoS_{2}$ can be used for tuning the tunneling phenomena. We study the effect of gate voltage on interaction free energy and Hamaker coefficient when $N=1$. Fig.5(a) and (b) show the variation of interaction free energy and Hamaker coefficient via $V_{g}$ for $T=300$ Kelvin $(K)$, respectively. As Eq.3 shows, for $V_{g}=0\rightarrow E_{F}=0$, but the term $ k_{B}T$ is not equal to zero in Eqs.4 and 5; therefore, the value of free energy differs from the related value shown in Fig.4(a) when $V_{g}=0$. This is the weakness point of the model. However according to the Figs. 5(a) and (b), the both interaction free energy and Hamaker coefficient decrease by applying $V_{g}<-50$ eV due to the decreasing the dielectric constant of $MoS_{2}$ layer. Therefore, one is able to tune the interaction free energy by applying an external voltage.
\begin{figure}
\includegraphics[width=0.8\linewidth]{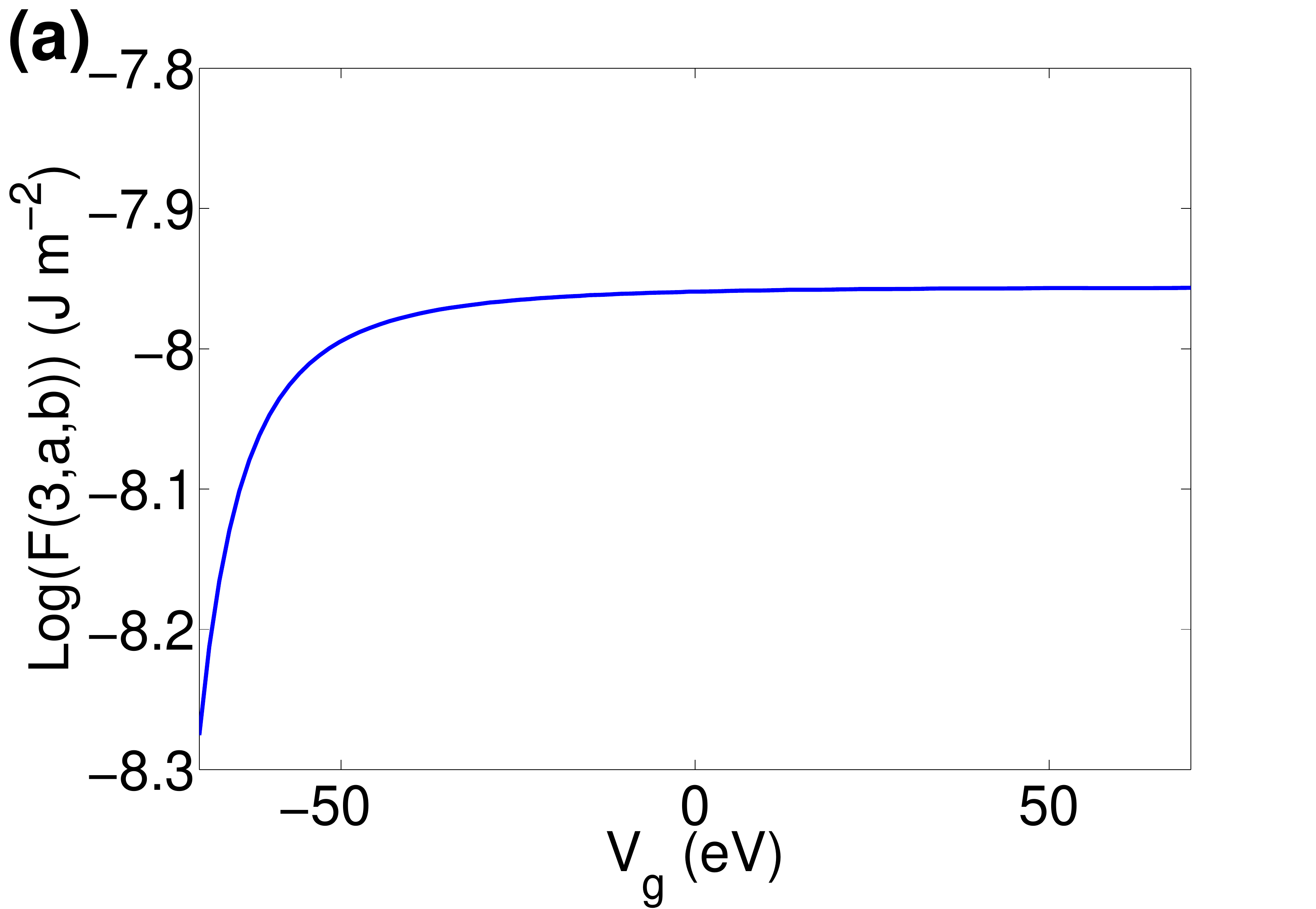}
\hspace{20pt}
\includegraphics[width=0.8\linewidth]{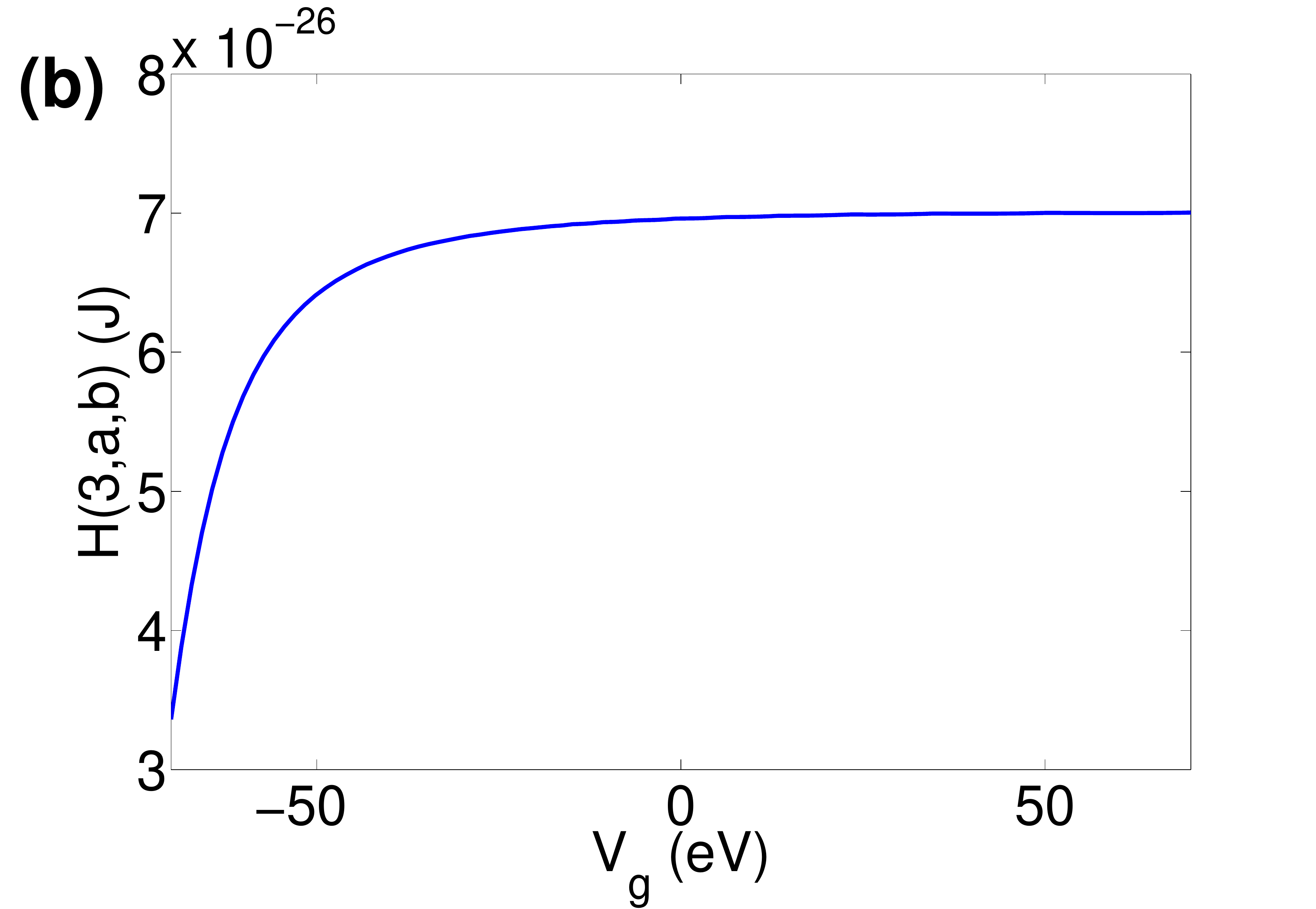}
\caption{(Color online) (a) The interaction free energy across one layer of $MoS_{2}$ and two layers of graphene and (b) Hamaker coefficient via gate voltage.}
\label{fig:Fig5}
\end{figure}
\subsection{Three-body configuration}
In this section, we consider a three-body configuration, shown in Fig.2, and study the evanescent photon tunneling between two silicon layers in presence and absence of the intermediate $MoS_{2}$ layer. We set the temperature of left side silicon at $T_{1}=1000 K$ and right side silicon at $T_{3}=300 K$. Then, we choose the temperature of the intermediate $MoS_{2}$ layer at $T_{2}$ such that the heat flux is amplified between left and right silicon layers. Therefore, for doing that, we use the three-body amplification model\cite{R17,R18} instead of the above stratified media model. We use the data of Ref.33 for dielectric constant of silicon\cite{R33} (see Fig.6).
\begin{figure}
\includegraphics{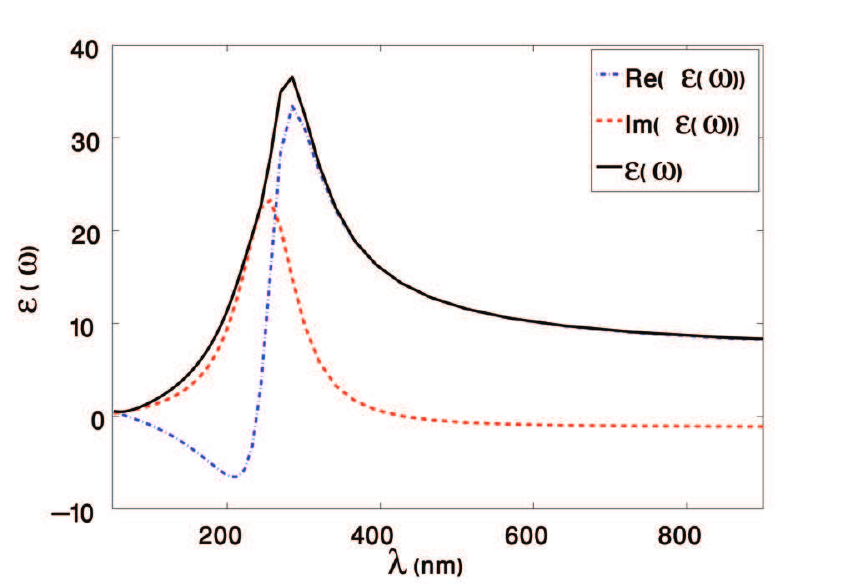}% Here is how to import EPS art
\caption{\label{fig:epsart}  (Color online) The dielectric constant of silicon.}
\end{figure}
Fig.7(a) shows the heat flux (in log-scale) between two silicon layers via distance between them $(d)$ and wavelength $(\lambda)$ when the temperature difference between layers is equal to $\Delta T=700 K$. As it shows, only for very small distance the heat flux has meaningful value. By adding the $MoS_{2}$ layer, coupling between modes happens and heat is amplified between silicon layers\cite{R17,R18}. As Fig.7(b) shows, for some specific values of $d$ and $\lambda$, the heat is amplified significantly. Now let us apply the gate voltage and see what is happened. Fig.8 shows the heat flux via wavelength when $d=10 nm$. As Fig.3(a) shows, three important peaks of the dielectric constant of $MoS_{2}$ are placed at 451, 619, and 666 nm. 
\begin{figure*}
\includegraphics[width=0.8\linewidth]{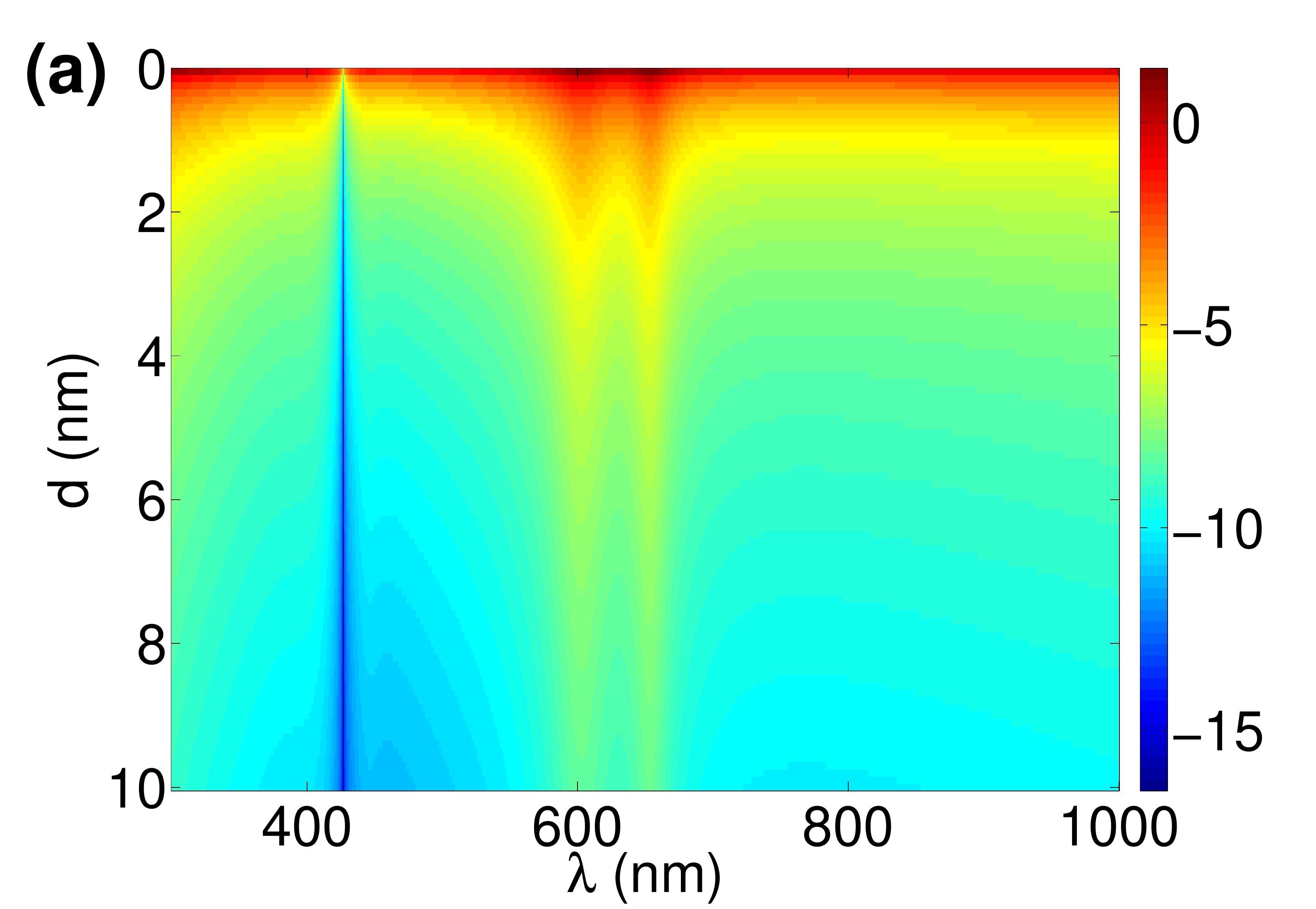}
\hspace{20pt}
\includegraphics[width=0.8\linewidth]{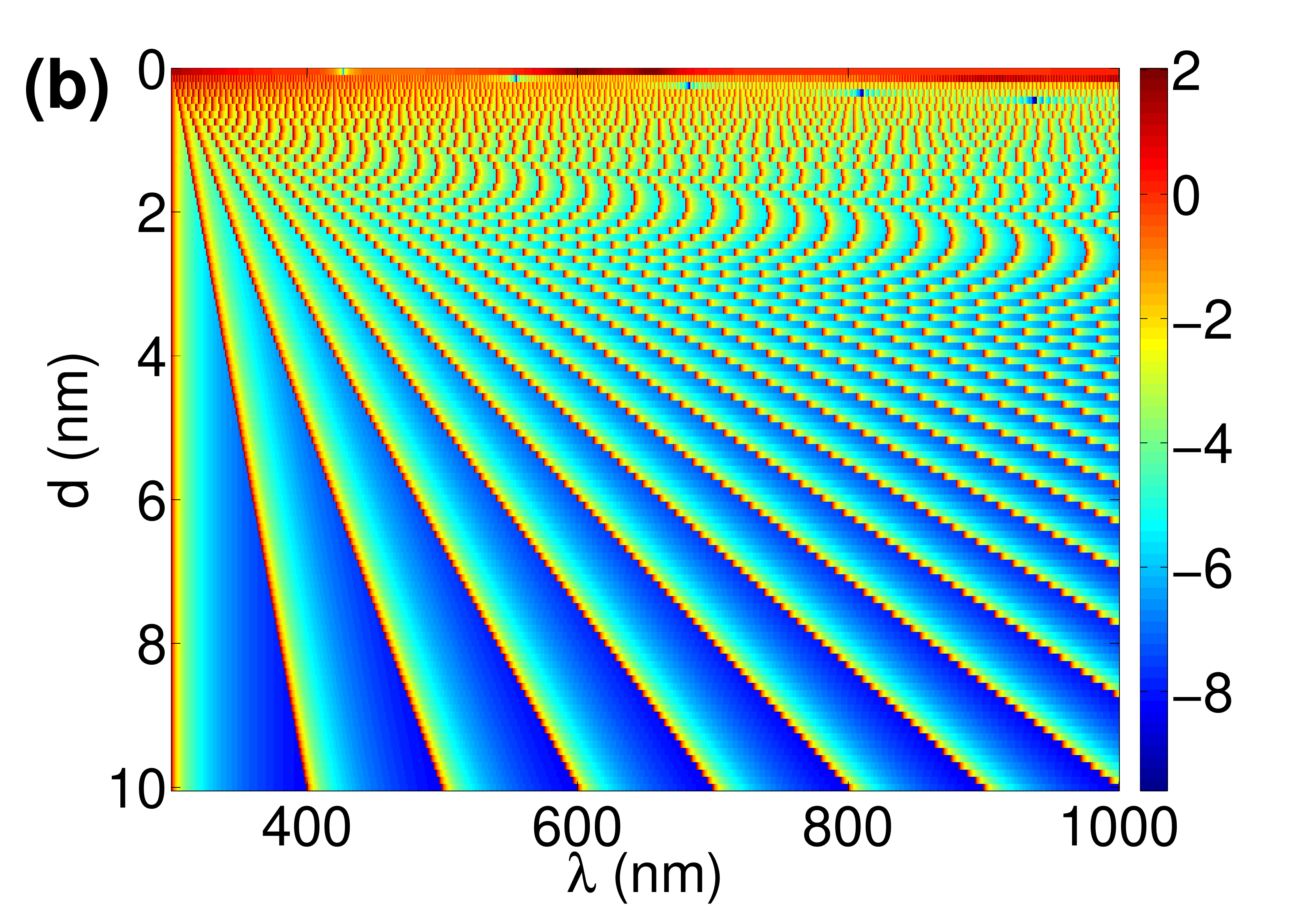}
\caption{(Color online) (a)  $Log(X^{*})$ via distance between two silicon layers, $d$, and wavelength $\lambda$, when $T_{1}=1000 K$ and $T_{2}=300 K$. (b) $Log(X^{*})$ via distance between each layer and intermediate $MoS_{2}$ layer, $d$, and wavelength $\lambda$, when $T_{1}=1000 K$ and $T_{3}=300 K$. The temperature of $MoS_{2}$ layer $(T_{2})$ should be tuned (see text).}
\label{fig:Fig7}
\end{figure*}
As Fig.8(a) shows, For $V_{g}=0$, two peaks of heat flux are placed at 602 and 654 nm and there is a valley at  425 nm, too. Thus, it can be concluded that, the coupling between modes is happened at peaks, while, the destructive coupling causes the reduction of heat flux at valley. By applying the gate voltage, the destructive coupling is amplified and the peaks of heat flux are disappeared (Fig.8(a)). According to the Fig.3(b), the peaks of $\varepsilon$ of $MoS_{2}$ are disappeared by applying the gate voltage. In consequence, in agreement with Fig.8(a), by changing $V_{g}$, the heat flux changes, and its peaks disappear. But, as Fig.8(b) shows, there are differences between heat fluxes for different values of $V_{g}$. Therefore, we can tune the heat flux by  applying the gate voltage. Since the silicon is used for manufacturing near infrared detectors, the technique can be used for increasing the sensibility of this kind of detectors similar to the medium wavelength infrared detectors\cite{R34}.  
\begin{figure*}
\includegraphics[width=0.8\linewidth]{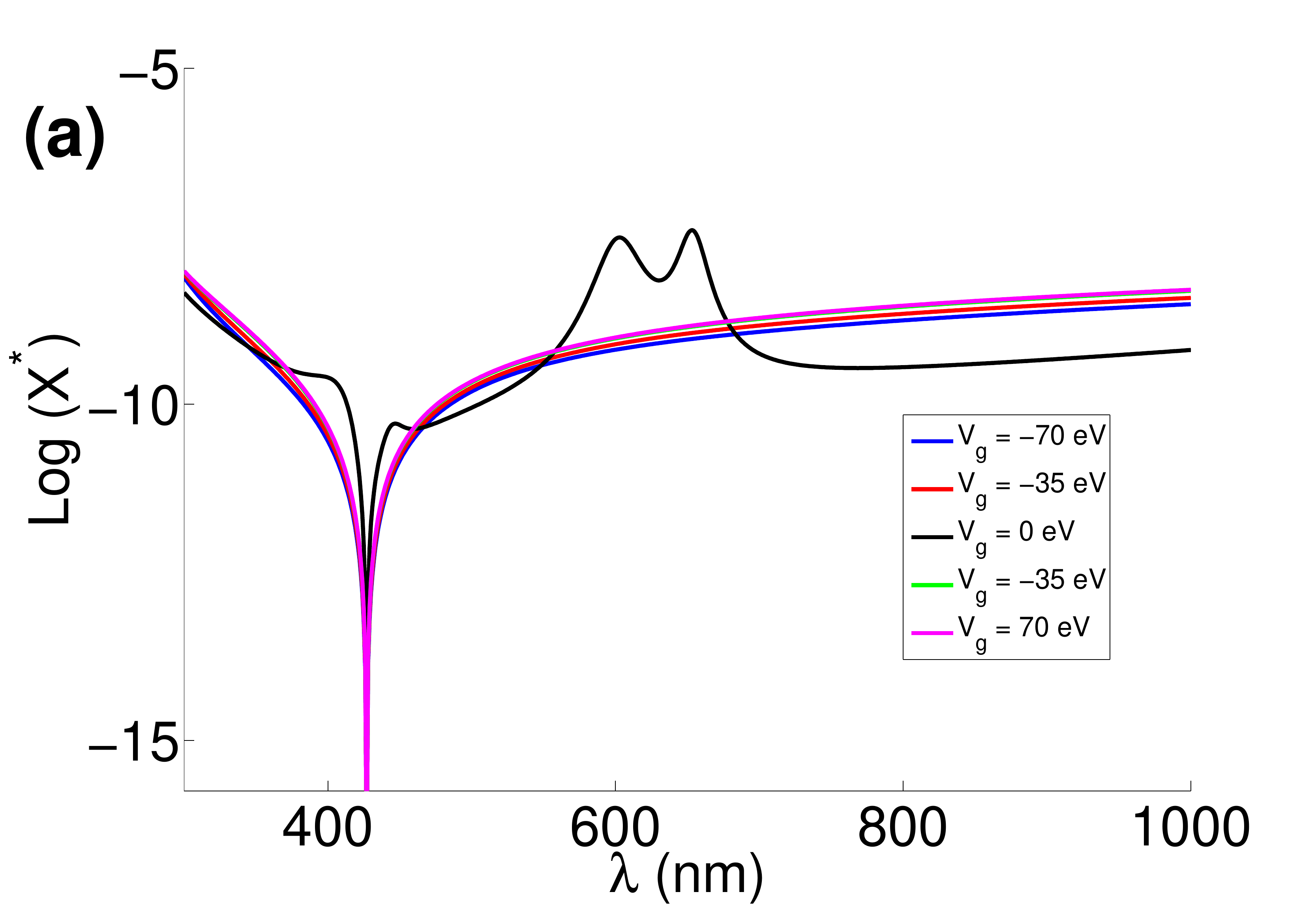}
\hspace{20pt}
\includegraphics[width=0.8\linewidth]{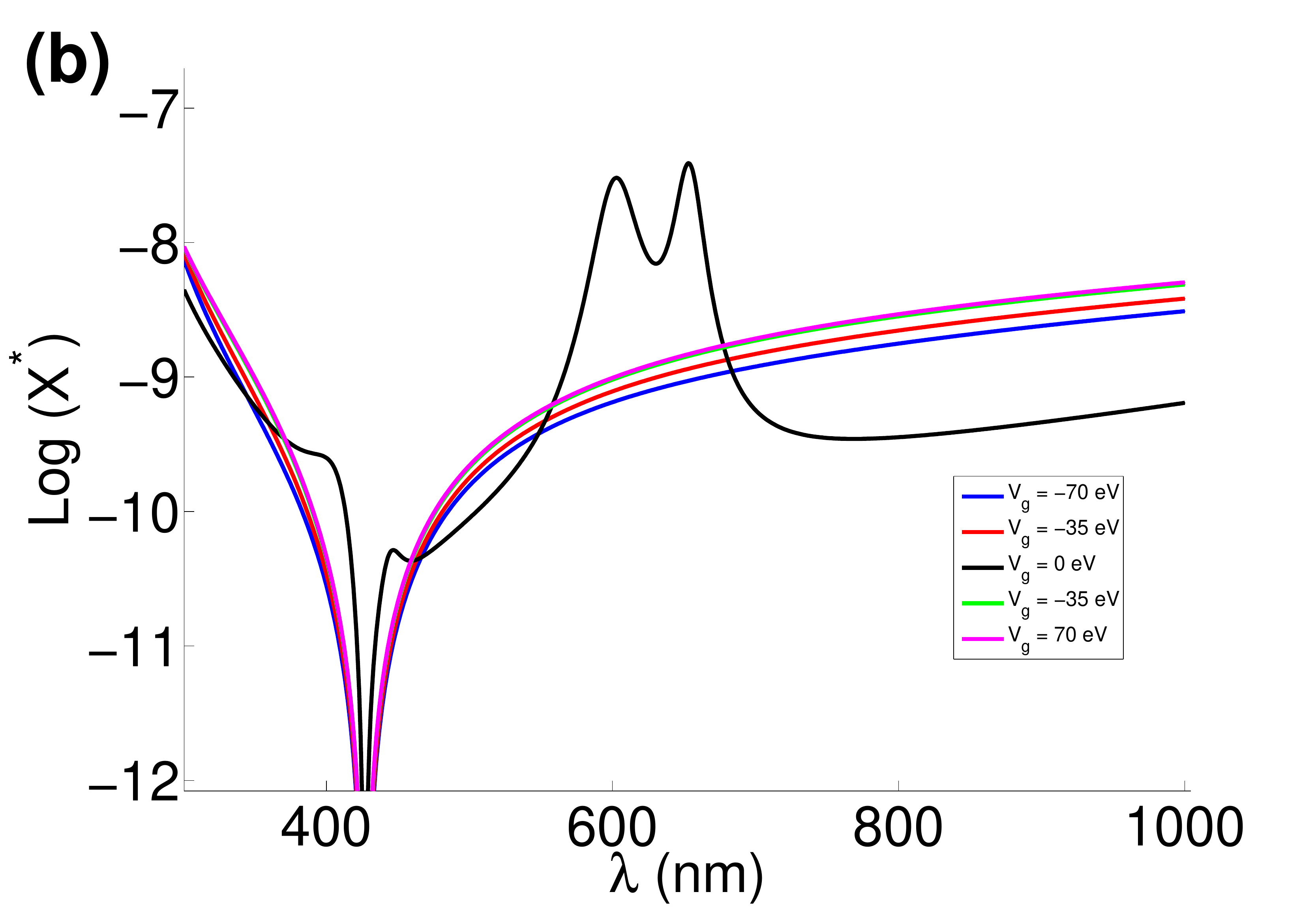}
\caption{(Color online) ) (a) $Log(X^{*})$ via wavelength $\lambda$, when the distance between silicon layer and $MoS_{2}$ layer, $d=10 nm$, and $T_{1}=1000 K$, $T_{3}=300 K$ and The temperature of $MoS_{2}$ layer $(T_{2})$ is tuned (see text). (b) insect of Fig.8(a).}
\label{fig:Fig8}
\end{figure*}
\section{Summary}
 We have considered a stratified media composed of graphene and $Mos_{2}$ layers and studied the interaction free energy in the media. We have shown that, when graphene is the nearest neighborhood of the leftmost and rightmost side of the media, the interaction free energy is higher than that of $Mos_{2}$ due to the lower thickness of graphene. Therefore, it could be considered as a rule. It has been shown that, the evanescent heat flux between two silicon layers is amplified if a monolayer of $MoS_{2}$ is placed between them. By applying a gate voltage to the intermediate $MoS_{2}$ layer, we have shown that, the heat flux between silicon layers could be tuned, and in consequence, the technique can be used for increasing the sensibility of near infrared detectors. We have only studied the monochromatic energy flux . Since the spectral properties vary significantly as a function of gate voltage, the frequency-intergrated flux could have interesting properties. The subject can be justified in next future studies.  

\nocite{*}

\bibliography{apssamp}% Produces the bibliography via BibTeX.

\end{document}